\documentclass[nofootinbib,twocolumn,preprintnumbers]{revtex4}
\usepackage[dvips]{graphicx}
\usepackage{amsmath,amsthm,amssymb}
\begin{document}

\def\lsim{\mathrel{\rlap{\lower4pt\hbox{\hskip1pt$\sim$}}
    \raise1pt\hbox{$<$}}}
\def\gsim{\mathrel{\rlap{\lower4pt\hbox{\hskip1pt$\sim$}}
    \raise1pt\hbox{$>$}}}
\newcommand{\vev}[1]{ \left\langle {#1} \right\rangle }
\newcommand{\bra}[1]{ \langle {#1} | }
\newcommand{\ket}[1]{ | {#1} \rangle }
\newcommand{\ev}{ {\rm eV} }
\newcommand{\kev}{{\rm keV}}
\newcommand{\mev}{{\rm MeV}}
\newcommand{\gev}{{\rm GeV}}
\newcommand{\tev}{{\rm TeV}}
\newcommand{\mpl}{$M_{Pl}$}
\newcommand{\mw}{$M_{W}$}
\newcommand{\Ft}{F_{T}}
\newcommand{\Zparity}{\mathbb{Z}_2}
\newcommand{\BLambda}{\boldsymbol{\lambda}}
\newcommand{\be}{\begin{eqnarray}}
\newcommand{\ee}{\end{eqnarray}}

\title{Searching for Higgs decays to four bottom quarks at LHCb}
\author{David E. Kaplan, Matthew McEvoy}
\affiliation{Department of Physics and Astronomy,
		Johns Hopkins University,
		Baltimore, MD  21218}
\date{\today}
\begin{abstract}
  We discuss the feasibility of seeing a Higgs boson which decays to four bottom quarks through a pair of (pseudo-)scalars at the LHCb experiment to argue that the use of $b$-physics triggers and off-line vertex reconstruction, as opposed to jet triggers with $b$ tagging, may be more effective for this signal.  Focusing on inclusive production for the Higgs, we find that for light scalar masses below 20 GeV,  signal reconstruction efficiencies of order a few percent may be enough for LHCb to find evidence for a Higgs with a dominant 4$b$ decay channel.
\end{abstract}

\maketitle

\section{Introduction}
In the last decade, tests of the electroweak model at accelerators give indirect evidence that the standard model is the correct description of physics below 1 TeV and that it contains a Higgs boson with a mass not too far above current limits \cite{lepewwg}.  The Higgs boson itself has not been seen \cite{Barate:2003sz}.  Assuming the standard model is correct for scattering energies beyond 1 TeV, we know to reasonable precision the production cross section, decay width and branching ratios for any given Higgs mass.  Thus search strategies can be optimized for such a Higgs.  However, Higgs phenomenology can be very sensitive to physics beyond the standard model.  The Higgs width, at a mass of 115 GeV, is about a thousandth of the $Z$ boson width.  Therefore, if there were neutral particles lighter than half the Higgs mass with any reasonable coupling to the Higgs, then the Higgs could decay dominantly into those particles, and the $Z$ width would not necessarily be significantly affected.

Here we assume that a decay into a pair of new light scalars dominates the Higgs width.  The scalar, $a$, subsequently decays to a pair of bottom quarks, producing the signal $h\rightarrow 4b$.  Such phenomenology has been discussed in the context of the NMSSM \cite{Gunion:1996fb,Stelzer:2006sp} and could also be possible in little Higgs or composite Higgs theories or any other such model where new light scalars (or pseudo-scalars) are coupled to the Higgs.

This signal has in fact been searched for at LEP II, with a bound on the mass of the Higgs of 110 GeV assuming a 100\% branching ratio to the light scalars and assume the scalars always decay to a pair of $b$ quarks, with weaker bounds for smaller branching ratios \cite{Abbiendi:2004ww}.  Searches at hadron colliders are considered very difficult due to the large irreducible background from QCD production of 4 $b$ jets.  It may be especially difficult at the ATLAS and CMS experiments \cite{AT,CMS} where triggers require high transverse momentum jets and/or leptons and thus the signal efficiency for what makes it to disk may be quite low.  The strategy usually suggested is thus to look at Higgs production in association with a $W$ boson decaying leptonically, even though the cross section is significantly lower than resonant production \cite{Carena:2007jk}.

In this note, we study the possibility of seeing the Higgs decay to 4 bottom quarks at the LHCb experiment \cite{LHCb}.  Its lower running luminosity allows for less stringent trigger requirements and its smaller solid angle coverage allows for a higher rate of events written to disk than other LHC experiments.  We find that for $a$ masses below 20 GeV there are simple kinematic cuts after which the signal approaches the size of the 4$b$ QCD background.  For these lighter $a$ masses, one is required to find two $b$ decay vertices within a single cone, or two almost merged jets.  The hope is that using non-standard vertexing techniques, such as topological vertexing \cite{petar} one may be able to seed calorimeter clusters with secondary vertices themselves with a reasonable efficiency.  In the next section, we estimate the total efficiency required, at different points in parameter space, to see evidence at 3 $\sigma$ for such a signal after one year of running (assuming 2 fb$^{-1}$ of data collected) at LHCb.

In principle, with the right triggers, ATLAS and CMS may do as well or better than LHCb for this signal.  We study LHCb because their planned inclusive muon trigger acts as an ``existence proof" that this data will be written to disk.  Also, in inclusive production, the Higgs is typically produced with a significant boost along the beam at the LHC, and thus at least one $b$ quark often lives at large pseudo-rapidity, and therefore excellent tracking and vertexing in the forward region may be crucial for this signal.  Finally, the possibility of doing useful energy frontier physics studies at the LHCb experiment could complement the ATLAS and CMS programs and should be explored.

To mock up the large number of theories which contain this physics, the `model' we are studying is simply the standard model with an additional real uncharged scalar $a$.  The new scalar can couple to the Higgs doublet via, for example,
\begin{equation}   {1\over 2}\lambda a^2 H^\dagger H .   \end{equation}
After electroweak symmetry breaking, this term will generate a tree-level mass squared for $a$ of $m_a^2 = (1/2) \lambda v^2$ and the trilinear coupling $(1/2) \lambda v h a^2$, where $v\simeq 246$ GeV is the Higgs' vacuum expectation value (vev).  The cubic coupling allows (for a light enough $a$) the Higgs to decay into a pair of scalars, with a rate
\begin{equation}
\Gamma(h\rightarrow aa) \simeq {1 \over 32\pi} \frac{\lambda^2 v^2}{m_h}\sqrt{1-{4 m_a^2 \over m_h^2}}
\end{equation}
When compared to the total decay rate of a ({\it e.g.}, 115 GeV) standard model Higgs, we see that for the above partial width to dominate the Higgs decay, the coupling to the scalar should be $\lambda>(.025)$.  If the physical mass of $a$ isn't a fine-tuned combination of different contributions ({\it i.e.}, a cancellation to no less than 10\%), then we should expect its mass to be above roughly 10 GeV if it is to dominate the Higgs width.

The most natural way for the singlet to decay is in the same way as a standard model Higgs boson of the same mass.  This would occur if the singlet mixes with the Higgs via a singlet vev or an operator $a |H|^2$.  The singlet could also be Higgs-like if it mixes with other Higgs-like states.  In the next-to-minimal supersymmetric standard model (NMSSM), a pseudo-scalar can be naturally light and mix with the heavier CP-odd part of the uneaten Higgs doublet.  In any case, even a tiny mixing of this type would allow a, say, 20 GeV scalar a prompt decay rate, dominantly into $b\bar{b}$.  We study precisely this signal setting the cross section of Higgs production to that of the standard model (at leading order) and the decay branching ratios of $h\rightarrow aa\rightarrow b\bar{b}b\bar{b}$ to unity.  The significances we quote scale with these quantities, for fixed backgrounds.  Other decay options can be generated with additional new physics at the electroweak scale.  See for example \cite{Gunion:1996fb}.  

\section{Signals and Backgrounds at LHCb}

The signal we studied was Higgs production with $h \rightarrow a a$ and $a \rightarrow b \bar{b}$. Signal events were produced and showered using Pythia v6.409 \cite{Sjostrand:2006za}.  Our analysis was performed for a number of different masses for $h$ and $a$. The study was done for $m_{h}$ at 115, 130 and 145 GeV with $m_{a}$ between 15 and 35 GeV in 5 GeV steps.  Pythia generates the signal at  leading order plus showering. At this order, the integrated cross-section for Higgs production at LHC energies is $\sim 25$pb for $m_h=115$ GeV, $\sim 20$pb for $m_h=130$ GeV, and $\sim 15$pb for $m_h=145$ GeV. Next-to-leading-order and higher corrections add $\sim 60$\% to this \cite{Kilgore:2002yw}, though we do not include the enhancement as we also run the background at leading order. We assume branching ratios of unity for both the $h \rightarrow aa$ and $a \rightarrow b \bar{b}$ decays.

The background we include in this study is QCD production of four $b$ quarks. QCD 2$b$ production does not contribute significantly after double-counting of events is taken into account. Background events were produced in ALPGEN v2.10 \cite{Mangano:2002ea} and then showered through Pythia v6.325.
Backgrounds not considered were those in which only three $b$ quarks or fewer are produced in the acceptance region (important when a $c$ or light quark or gluon is mistagged as a $b$ quark) \footnote{We are counting partons as opposed to jets as an important signal region occurs when pairs of $b$ jets are partially merged.}.  We do not study these channels as their importance depends sensitively on mistag rates, and using estimates of jet mistagging at the Tevatron suggests that these contributions will not dominate the background \cite{Stelzer:2006sp}. 

The $b\bar{b}$ QCD background is potentially more significant in the kinematic region where the pairs of $b$ jets from the individual scalar decays merge ({\it i.e.}, when the scalars are light).  QCD production of two $b$ jets will be enormous at the LHC, with a cross section on the order of 1 mb -- roughly a factor of 1000 larger than 4$b$ production.  Thus the technique for tagging two merged $b$ jets better include a very small 1 $b$ to 2 $b$ fake rate (less than a few percent).

It is expected that signal and background events will pass the level zero (L0) trigger at LHCb with high efficiency, while the recent upgrade of the LHCb read out rate to $2$ kHz has permitted the addition of an inclusive high-level muon trigger. The inclusive branching ratio of $b$ hadrons to muons is $\sim 11\%$, so given 4 $b$'s per event, there is a $\sim 36\%$ probability that a given event will see the decay of at least one $b$ hadron to a muon. The $b$'s from signal events are expected to be fairly hard for the forward acceptance range of LHCb $p_T$ ($\gsim$ $10$ GeV) given the large invariant mass of the Higgs boson, so it seems reasonable to assume that  $>50\%$ of these muon decays will be triggered on. This trigger path by itself therefore gives an expected trigger efficiency of order or better than $20\%$ \cite{nakada}. Other high-level triggers which are already in place or could be designed for this purpose may increase this efficiency significantly and are worthy of study.

To avoid generic QCD backgrounds, we require the tagging of all four $b$ quarks.  However, rather than first reconstructing jets and then applying a b-tag (the usual method), we assume the four $b$ vertices are found first using a method such as 'topological vertexing' \cite{petar}.  Then, the $b$-hadrons are used as seeds of jets with cone-size $R=\sqrt{\delta\phi^2 + \delta\eta^2}=0.6$ (where $\delta\eta$ and $\delta\phi$ are the spread in pseudo rapidity and azimuthal angle).  When the jets overlap, we divide up the momenta of particles in the overlapping region between the two jets based on a smooth function of distance from the seeds, though in practice this detail is unimportant; pairs of overlapping jets are reconstructed into scalars, and only information about these reconstructed scalars is used as a kinematic discriminant.  As a rough approximation, we take the triggering and tagging efficiencies as equal for signal and background and independent of kinematic configuration.  We define $\Sigma \equiv \Sigma_0/(\sqrt{\epsilon_{tag}\epsilon_{trig}})$ where $\Sigma$ is the significance, $\epsilon_{tag}$ is the efficiency to tag all four $b$ jets,  $\epsilon_{trig}$ is the trigger efficiency and $\Sigma_0$ is the significance (described below) assuming $\epsilon_{trig} = \epsilon_{tag} = 1$. Our results are described as an estimate the efficiencies required to generate a significance $\Sigma$ of $3.0\sigma$.

Beyond $b$-tagging, we require all $b$ jets to have $p_T>10$ GeV, and of course for all of them to live within the detector acceptance.  LHCb's nominal acceptance range in pseudorapidity $1.9\leq \eta \leq 4.3$ on the vertical axis and $2.1\leq\eta\leq 4.3$ on the horizontal axis.

Once these four $b$ jets have been reconstructed, there are three possible ways to combine them into two pairs of jets in order to reconstruct the two $a$ particles.  Requiring the two pairs to have the same invariant mass within detector resolution turns out not to be a useful discriminant.  The combination which is chosen is that which minimizes the quantity $(R_1)^2 + (R_2)^2$ where $R_1$ is the previously-defined distance in $\eta$-$\phi$ space between the first pair of jets and $R_2$ is the distance between the second pair. It should be noted that this technique limits us to the study of relatively light scalars whose decay products tend to be collimated. Our method chooses the correct combination of jets $\geq 90$\% of the time for all the previously mentioned sets of parameters other than with $m_h=130$ GeV, $m_a=35$ GeV and $m_h=115$ GeV, $m_a=35$ GeV where the proper combination was recreated $88$\% and $76$\% of the time, respectively. For scalar masses closer to the upper kinematic limit ($m_a\sim m_h/2$) it is possible to recreate the proper jet combination moderately well (up to $\sim 70$\% when $m_a\gsim0.48m_h$) by choosing the combination which maximizes $(R_1)^2 + (R_2)^2$. Unfortunately, this method does not significantly reduce the background once an overall invariant mass cut is placed on the four jets.

\begin{figure}
\resizebox{\hsize}{!}{\includegraphics{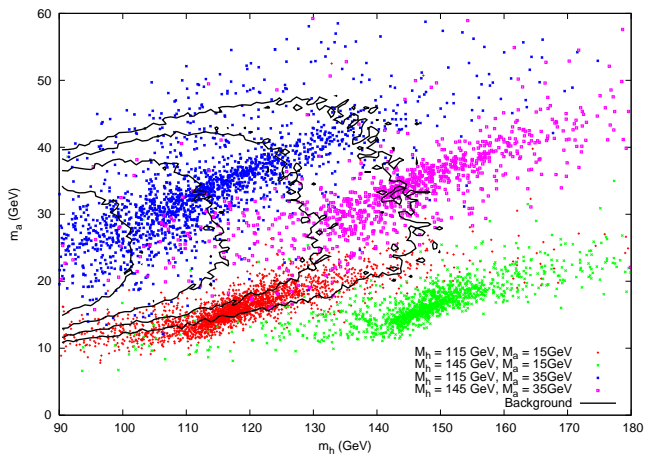}}
\caption{Reconstructed scalar mass versus reconstructed Higgs mass for $2$ fb$^{-1}$ worth of signal at different points in parameter space along with contour lines for background. Each contour (moving from left to right) represents a halving of the background intensity, with the outermost contour at a background density (in $m_a$ vs. $m_h$ space) of $\sim 128$ events/GeV$^2$ given $2$ fb$^{-1}$ of data. Note that the characteristic slope of the signal distributions is relatively constant across a large region of parameter space.}
\label{fig:scatterplot}
\end{figure}

From each event, we have therefore reconstructed a Higgs mass and two $a$ masses. The mean of these two $a$ masses is taken as the $a$ mass in all further analyses. When $m_h$ is plotted against $m_a$ as in Figure ~\ref{fig:scatterplot} signal events generated by different sets of parameters tend to group into elongated ellipses. Events generated with a lower $m_a$ tend to be more tightly grouped than those with a higher mass, but the slope of the long axis of the ellipse seems to be independent of the location in parameter space and is $\sim\frac{1}{3}$. See the appendix for details.  This spread in invariant masses is a result of the jet (cone) algorithm we have chosen but does not take into account general smearing due to detector effects.  It would be interesting to see if a nearly constant slope is maintained in a full simulation.

The fact that this slope is fairly constant across parameter space allows us to perform the following transformation:
\begin{align}
&q^- = m_h - 3m_a \nonumber \\
&q^+ = m_h + 0.333m_a
\label{eq:reparam}
\end{align}
The resulting quantities $q^-$ and $q^+$ are therefore roughly uncorrelated for a set of signal events from any point in a large region of parameter space. The standard deviation of $q^-$ for such a set is of the order of $5-10$ GeV, while that of $q^+$ is of the order of $20$ GeV.

\begin{table}[b]
\caption{Efficiency required, $\epsilon_{req}$ to see a $3.0\sigma$ significance with 2 fb$^{-1}$ of data for various values of $m_h$ and $m_a$ along with the point (in $q^+$ and $q^-$) at which the greatest significance is seen. The ratio of signal to background, $S/B$, included in this significance is also provided.  The first four columns are in units of GeV.  The required efficiency scales approximately with the inverse square root of the total luminosity.}
\begin{tabular}{||c|c|c|c|c|c||}
\hline
\: $m_h$ \:      & \: $m_a$ \: &\; $q^+$ \; & $q^-$ & \; $\epsilon_{req}$ \; &\; $S/B$ \; \\ \hline \hline
$115$&$15$&$105$&$72.5$&$0.06$&$0.11$\\ \hline
$115$&$20$&$135$&$57.5$&$0.24$&$0.023$\\ \hline
$115$&$25$&$135$&$42.5$&$0.39$&$0.016$\\ \hline
$115$&$30$&$135$&$27.5$&$0.69$&$0.012$\\ \hline
$115$&$35$&$135$&$12.5$&$1.15$&$0.009$\\ \hline
$130$&$15$&$125$&$87.5$&$0.05$&$0.175$\\ \hline
$130$&$20$&$145$&$72.5$&$0.24$&$0.034$\\ \hline
$130$&$25$&$155$&$57.5$&$0.39$&$0.025$\\ \hline
$130$&$30$&$155$&$42.5$&$0.59$&$0.020$\\ \hline
$130$&$35$&$155$&$27.5$&$0.88$&$0.017$\\ \hline
$145$&$15$&$135$&$102.5$&$0.05$&$0.38$\\ \hline
$145$&$20$&$155$&$87.5$&$0.22$&$0.052$\\ \hline
$145$&$25$&$165$&$72.5$&$0.46$&$0.029$\\ \hline
$145$&$30$&$165$&$57.5$&$0.78$&$0.022$\\ \hline
$145$&$35$&$165$&$42.5$&$1.00$&$0.020$\\ \hline
\end{tabular}
\label{tab:signif}
\end{table}

To proceed, we center a box of size $15$ GeV in $q^-$ and $50$ GeV in $q^+$ at various points $(a_i,b_j)$ on a grid in $(q^-,q^+)$ space and count the number of signal and background events falling in each of these boxes. These numbers are used to estimate the significance for the central point of each box where the significance $\Sigma_0 \equiv S/ \sqrt{S+B}$, where $S$ and $B$ are the number of signal and background points in the box respectively.  Adjacent central points are separated by $a_{i+1}-a_i = 5$ GeV and $b_{j+1}-b_j = 10$ GeV. The boxes around neighbouring central points overlap, therefore their significances are correlated and cannot be independently combined. No trials factor is included. Table \ref{tab:signif} provides a list of the points with the highest significances for various pairs of parameters $(m_a, m_h)$. The significance of each of these points is presented as the efficiency ($\epsilon_{trigg}*\epsilon_{tag}$) required to see a significance of $3.0\sigma$ with $2$ fb$^{-1}$ (as an approximation, taking both signal and background to have identical efficiencies and that these efficiencies are independent of kinematic configurations). It should be noted that in all cases the $15$ GeV by $50$ GeV box with the highest significance captured a large fraction of signal events (ranging from 0.4 to 0.7, with a negative correlation between $m_a$ and the capture rate). Interestingly, for cases where $m_a$ was $15$ GeV the highest significance was seen at $q^+$ slightly lower than the signal peak. This is a result of the background dropping much faster toward lower $q^+$ in this region than the signal, also giving a significantly higher signal-to-background ratio. It is evident from these results that the difficulty of detecting this decay of the Higgs is fairly independent of the mass of the Higgs boson, while it is strongly dependent on the scalar mass. 

\begin{figure}[t]
\resizebox{\hsize}{!}{\includegraphics{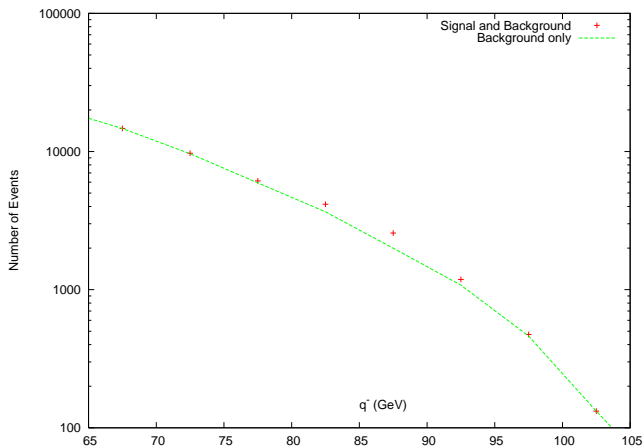}}
\caption{Number of events versus $q^-$ for $2$ fb$^{-1}$ worth of signal and background restricted to events with $q^+$ in the range $100$ GeV $\leq q^+ \leq 150$ GeV assuming $m_h = 130$ GeV and $m_a = 15$ GeV. The range of $q^+$ represents the region of maximum significance for this signal.}
\label{fig:sigbck}
\end{figure}

We present the results for one point in parameter space ($m_h = 130$ GeV, $m_a = 15$ GeV) graphically in Figure~\ref{fig:sigbck} by restricting ourselves to events with $100$ GeV $\leq q^+ \leq 150$ GeV and histogramming these events by $q^-$ into bins of size $5$ GeV. The range of $q^+$ is that which gives the maximum significance for these parameters. 

\section{Conclusion: Estimated Reach}

From the data it is apparent that given a relatively light pseudo-scalar and robust $b$-tagging and triggering efficiencies the reach of this search technique extends across most or all Higgs masses where decays into pseudo-scalars could dominate (from the lower bound on the Higgs mass of $\sim 115$ GeV to somewhat less than twice the Z-mass). It is also apparent that for pseudo-scalar masses greater than $~30$ GeV detection of the Higgs using our technique is impossible. Logarithmically, the region of detectability $2m_b$ $\leq m_a \leq 30$ GeV represents slightly more than half the range of masses where decays of Higgs bosons to $b$ quarks could be expected to dominate ($2m_b$ $\leq m_a \leq \frac{m_h}{2}$. Achieving this reach is heavily dependent on efficient and accurate $b$-tagging, especially in configurations where two $b$ jets overlap. 

\vskip 0.1 cm

We thank Aurelio Bay, Petar Maksimovic, Mark Mathis, Olivier Schneider,
and Frederic Teubert for useful discussions.
This work is supported in part by NSF grants PHY-0244990 and PHY-0401513,
by DOE grant DE-FG02-03ER4127, and by the Alfred P. Sloan Foundation.

\section{Appendix: Correlation of errors in the measurement of scalar and Higgs masses}

Let the first scalar have momentum $p^\mu$ and the second scalar have momentum $q^\mu$ in the rest frame of the higgs boson, while the error in measurement of each of these 4-vectors is $\delta^\mu(p)$ and $\delta^\mu(q)$ respectively. The errors in measurement of the scalar masses are:

\begin{align}
&\Delta m_{a_1} = [(p^\mu+\delta^\mu(p))(p_\mu=\delta_\mu(p))]^{\frac{1}{2}}-[p^\mu p_\mu]^{\frac{1}{2}} \nonumber \\
&\Delta m_{a_2} = [(q^\mu+\delta^\mu(q))(q_\mu+\delta_\mu(q))]^{\frac{1}{2}}-[q^\mu q_\mu]^{\frac{1}{2}}
\label{eq:dma1}
\end{align}

Assuming that $p^\mu p_\mu \gg p^\mu \delta_\mu(p) \gg \delta^\mu(p) \delta_\mu(p)$ we drop terms containing factors of $\delta^\mu(p) \delta_\mu(p)$ and approximate the square root to first order in $\frac{p^\mu\delta_\mu(p)}{p^\mu p_\mu}$. We carry out the same operation on the second scalar as well, remembering that $p^\mu p_\mu = q^\mu q_\mu = (m_a)^2$:

\begin{align}
&\Delta m_{a_1} \simeq \frac{p^\mu\delta_\mu(p)}{m_{a}} \nonumber \\
&\Delta m_{a_2} \simeq \frac{q^\mu\delta_\mu(q)}{m_{a}}
\label{eq:dma2}
\end{align}

Boosting to the rest frames of the respective pseudo-scalars,

\begin{align}
&\Delta m_{a_1} \simeq \frac{p^{'\mu}\delta^{'}_\mu (p)}{m_{a}} = \delta^{'}_0(p) \nonumber \\
&\Delta m_{a_2} \simeq \frac{q^{''\mu}\delta^{''}_\mu (q)}{m_{a}} = \delta^{''}_0(q)
\label{eq:dma3}
\end{align}

Remembering our definition of the measured scalar mass $m_a(meas) \equiv \frac{m_{a_1}(meas) + m_{a_2}(meas)}{2}$,

\begin{equation}
\Delta m_a = \frac{\Delta m_{a_1} + \Delta m_{a_2}}{2} \simeq \frac{\delta^{'}_0(p) + \delta^{''}_0(q)}{2}
\label{eq:dma4}
\end{equation}

We now examine the error in the measurement of the Higgs mass:

\begin{align}
\Delta m_h = &[(p^\mu + q^\mu + \delta^\mu(p) + \delta^\mu(q))&\nonumber \\
&(p_\mu + q_\mu + \delta_\mu(p) + \delta_\mu(q))]^{\frac{1}{2}}&\nonumber \\
& - [(p^\mu + q^\mu)(p_\mu + q_\mu)]^{\frac{1}{2}}&
\label{eq:dmh1}
\end{align}

We again drop terms containing $\delta^2$ and approximate the square root to first order in $\frac{(p^\mu + q^\mu)(\delta_\mu(p) + \delta_\mu(q))}{(p^\mu + q^\mu)(p_\mu + q_\mu)}$, remembering that $(p^\mu + q^\mu)(p_\mu + q_\mu) = (m_h)^2$ and that we are in the rest frame of $p^\mu + q^\mu$:

\begin{equation}
\Delta m_h \simeq \frac{(p^\mu + q^\mu)(\delta_\mu(p) + \delta_\mu(q))}{m_h} = \delta_0(p) + \delta_0(q)
\label{eq:dmh2}
\end{equation}

We now make the assumption that the major sources of error in the measurement of the masses derive from ultra-relativistic particles originating from the Higgs which escape the two cones around the component $b$ hadrons of each scalar and other ultra-relativistic particles originating in the underlying event which happen to enter these cones.

\begin{align}
& = \delta_{\mu}(p) = \sum_i \tau_{i_\mu}(p) \nonumber-\sum_j \rho_{j_\mu}(p) \nonumber \\
& = \delta_{\mu}(q) = \sum_k \tau_{k_\mu}(q) \nonumber-\sum_l \rho_{l_\mu}(q) \nonumber \\
& = \delta^{'}_{\mu}(p) = \sum_i \tau^{'}_{i_\mu}(p) \nonumber-\sum_j \rho^{'}_{j_\mu}(p) \nonumber \\
& = \delta^{''}_{\mu}(q) = \sum_k \tau^{''}_{k_\mu}(q) - \sum_l \rho^{''}_{l_\mu}(q)
\label{eq:sumEs1}
\end{align}

Given particle $i$ from the underlying event entering the jet cone of $a_1$, while $\vec{\tau_i^{'}}(p)$ makes an angle of $\phi^{'}_i$ with $\vec{p}$ and $a_1$ has a boost of $\gamma_{a_1}$ and velocity $\beta_{a_1}$ relative to the Higgs, then

\begin{align}
\tau_{i_0}(p) & = \gamma_{a_1}\tau_{i_0}^{'}(p)(1 + \beta_{a_1}\cos{\phi_i^{'}}) \nonumber \\
& = \frac{m_h}{2m_a} \biggl( 1 + \sqrt{1 - \Bigl(\frac{2m_a}{m_h}\Bigr)^2}\cos{\phi_i^{'}} \biggr) \tau_{i_0}^{'}(p) \nonumber \\
& = A(\phi_i^{'}) \frac{\tau_{i_0}^{'}(p)}{2}
\label{eq:Erat1}
\end{align}

Similarly, given particle $j$ (originating from $a_1$) escaping the two relevant jet cones, while $\vec{\rho_j^{'}}(p)$ makes an angle of $\theta^{'}_j$ with $\vec{p}$ and $a_1$ has a boost of $\gamma_{a_1}$ and velocity $\beta_{a_1}$ relative to the Higgs, then

\begin{align}
\rho_{j_0}(p) & = \gamma_{a_1}\rho_{j_0}^{'}(p)(1 + \beta_{a_1}\cos{\theta_j^{'}}) \nonumber \\
& = \frac{m_h}{2m_a} \biggl( 1 + \sqrt{1 - \Bigl(\frac{2m_a}{m_h}\Bigr)^2}\cos{\theta_j^{'}} \biggr) \rho_{j_0}^{'}(p) \nonumber \\
& = B(\theta_i^{'}) \frac{\rho_{i_0}^{'}(p)}{2}
\label{eq:Erat2}
\end{align}

In the frame where $a_1$ has no momentum in the $z$ direction and the Higgs has no transverse momentum (the ``transverse frame''), given an ultra-relativistic particle making an angle of $\alpha^T$ with $\vec{p^T}$ (the transverse component of $\vec{p}$) it makes, in the rest frame of $a_1$ an angle of $\alpha^{'}$ with $\vec{p^T}$ and

\begin{equation}
\cos{\alpha^{'}} = \frac{\cos{\alpha^T}+\beta^T_{a_1}}{1+\beta^T_{a_1}\cos{\alpha^T}}
\label{eq:costrans}
\end{equation}

where $\beta^T_{a_1}$ is the boost of $a_1$ in the transverse frame. If the Higgs has no transverse momentum and $\vec{p}$ makes an angle $\Psi$ with the beam axis, then

\begin{equation}
\beta^T_{a_1} = \frac{\beta_{a_1}\sin{\Psi}}{1-\beta_{a_1}^2\cos^2{\Psi}}
\label{eq:betat}
\end{equation}

We model the union of the two $b$ hadron cones as a single cone of radius $R_a$ (in $\eta$-$\phi$ space) centered on the scalar. In the transverse frame, this cone is approximately a cone in cartesian space with $\alpha^T_{max} = R_a$.

Now, the particles from the underlying event are approximately uniformly distributed over $\eta$, so in the transverse frame they are isotropically distributed over small $\alpha^T$ (since $\Delta\eta$ is relatively insensitive to boost along the beam axis and $\eta \simeq (\pi / 2 - \theta)$ for $\pi/2 - \theta$ small). The particles from the underlying event which contribute to measurement error fall into the jet cone. We therefore use eqns. \ref{eq:Erat1}, \ref{eq:costrans} and \ref{eq:betat} to write $A(\phi^{'})$ in terms of $\phi^T$ (the angle between the particle from the underlying event and the scalar in the transverse frame) and $\Psi$ (the angle between the scalar and the beam axis in the rest frame of the Higgs), averaging over $0 \leq \phi^T \leq R_a$ and $0 \leq \Psi \leq \pi$, using

\begin{equation}
\overline{A} = \frac{\int_0^\pi d\Psi \int_0^{R_a} d\phi^T A(\phi^T, \Psi) \sin{\Psi} \sin{\phi^T}}{\int_0^\pi d\Psi \int_0^{R_a} d\phi^T \sin{\Psi} \sin{\phi^T}}
\label{eq:Aave}
\end{equation}

Meanwhile, daughter particles of $a_1$ are (over an ensemble of decays) distributed isotropically in the reference frame of $a_1$. In this frame, the jet covers an area from $\alpha^{'} = 0$ to $\alpha^{'}_{max} = \alpha^{'}_{max}(\alpha^T_{max}, \Psi) = \alpha^{'}_{max}(R_a, \Psi)$ using eqns. \ref{eq:costrans} and \ref{eq:betat}. The daughter particles which contribute to measurement error are those which fall outside this cone. We average over $\alpha^{'}_{max}(R_a, \Psi) \leq \theta^{'} \leq \pi$ and $0 \leq \Psi \leq \pi$:

\begin{equation}
\overline{B} = \frac{\int_0^\pi d\Psi \int_{\alpha^{'}_{max}}^{\pi} d\theta^{'} B(\theta^{'}) \sin{\Psi} \sin{\theta^{'}}}{\int_0^\pi d\Psi \int_{\alpha^{'}_{max}}^{\pi} d\theta^{'} \sin{\Psi} \sin{\theta^{'}}}
\label{eq:Bave}
\end{equation}

\begin{table}
\caption{$\overline{A}$ and $\overline{B}$ using $R_a = 0.85$ for various points in parameter space.}
\begin{tabular}{|c|c|c|}
\hline
$\frac{m_h}{m_a}$&$\overline{A}$&$\overline{B}$\\
\hline
$4.0$&$4.6$&$1.8$\\
\hline
$5.0$&$4.9$&$1.7$\\
\hline
$6.0$&$5.1$&$1.5$\\
\hline
$7.0$&$5.2$&$1.4$\\
\hline
$8.0$&$5.2$&$1.3$\\
\hline
$9.0$&$5.2$&$1.3$\\
\hline
\end{tabular}
\label{tab:ratio}
\end{table}

Finally, we note that the above calculations hold true for mis-measurements of the second pseudo-scalar's 4-momentum and their effects on the measurement of $m_a$ and $m_h$.

If the error in the measurement in $m_a$ was entirely due to particles from the underlying event then we could approximate $\Delta m_h \simeq \Delta m_a \overline{A}$. If it was entirely due to daughter particles of the scalars escaping the jet cones then $\Delta m_h \simeq \Delta m_a \overline{B}$. Table \ref{tab:ratio} gives values for $\overline{A}$ and $\overline{B}$ for different mass ratios $\frac{m_h}{m_a}$ assuming $R_a = 0.85$ (slightly larger than the individual cone size). We can expect $\frac{\Delta m_h}{\Delta m_a}$ to lie somewhere between these two extremes. Note that $\overline{A} > 3.0$ while $\overline{B} < 3.0$.


\begin{thebibliography}{1}

\bibitem{lepewwg}
For a summary of the standard model fit to LEP, SLC and Tevatron data, see http://lepewwg.web.cern.ch/LEPEWWG/.

\bibitem{Barate:2003sz}
  R.~Barate {\it et al.}  [LEP Working Group for Higgs boson searches],
  Phys.\ Lett.\  B {\bf 565}, 61 (2003)
  [arXiv:hep-ex/0306033].


\bibitem{Gunion:1996fb}
 J.~F.~Gunion, H.~E.~Haber and T.~Moroi,
  eConf {\bf C960625}, LTH095 (1996);
  [arXiv:hep-ph/9610337];
R.~Dermisek and J.~F.~Gunion,
  Phys.\ Rev.\ Lett.\  {\bf 95}, 041801 (2005)
  [arXiv:hep-ph/0502105];
  S.~Chang, P.~J.~Fox and N.~Weiner,
  arXiv:hep-ph/0511250.
  \bibitem{Graham:2006tr}
  P.~W.~Graham, A.~Pierce and J.~G.~Wacker,
  arXiv:hep-ph/0605162.

\bibitem{Stelzer:2006sp}
  T.~Stelzer, S.~Wiesenfeldt and S.~Willenbrock,
  Phys.\ Rev.\  D {\bf 75}, 077701 (2007)
  [arXiv:hep-ph/0611242].

\bibitem{Abbiendi:2004ww}
  G.~Abbiendi {\it et al.}  [OPAL Collaboration],
  Eur.\ Phys.\ J.\ C {\bf 37}, 49 (2004)
  [arXiv:hep-ex/0406057];
  J.~Abdallah {\it et al.}  [DELPHI Collaboration],
  Eur.\ Phys.\ J.\ C {\bf 38}, 1 (2004)
  [arXiv:hep-ex/0410017].


\bibitem{AT}
{\it http://atlas.web.cern.ch/Atlas/index.html}
\bibitem{CMS}
{\it http://cms.cern.ch/}


\bibitem{Carena:2007jk}
  M.~Carena, T.~Han, G.~Y.~Huang and C.~E.~M.~Wagner,
  JHEP {\bf 0804}, 092 (2008)
  [arXiv:0712.2466 [hep-ph]].

\bibitem{LHCb}
{\it http://lhcb.web.cern.ch/lhcb/}


\bibitem{Sjostrand:2006za}
  T.~Sjostrand, S.~Mrenna and P.~Skands,
  JHEP {\bf 0605}, 026 (2006)
  [arXiv:hep-ph/0603175].

\bibitem{Kilgore:2002yw}
  W.~B.~Kilgore and R.~V.~Harlander,
  arXiv:hep-ph/0205152.

\bibitem{Mangano:2002ea}
  M.~L.~Mangano, M.~Moretti, F.~Piccinini, R.~Pittau and A.~D.~Polosa,
  JHEP {\bf 0307}, 001 (2003)
  [arXiv:hep-ph/0206293].






\bibitem{Dimopoulos:1981zb}
  S.~Dimopoulos and H.~Georgi,
  Nucl.\ Phys.\  B {\bf 193}, 150 (1981).
For a review, see
  S.~P.~Martin,
  arXiv:hep-ph/9709356.

\bibitem{Giudice:2006sn}
  See, for example, G.~F.~Giudice and R.~Rattazzi,
  Nucl.\ Phys.\  B {\bf 757}, 19 (2006)
  [arXiv:hep-ph/0606105].
\bibitem{D0qg} The D0 Collaboration,
D0 Note 5312 (2007).

\bibitem{Schael:2006cr}
  S.~Schael {\it et al.}  [ALEPH Collaboration],
  Eur.\ Phys.\ J.\  C {\bf 47}, 547 (2006)
  [arXiv:hep-ex/0602042].

\bibitem{Achard:2001ek}
  P.~Achard {\it et al.}  [L3 Collaboration],
  Phys.\ Lett.\ B {\bf 524}, 65 (2002)
  [arXiv:hep-ex/0110057];
  A.~Heister {\it et al.}  [ALEPH Collaboration],
  Eur.\ Phys.\ J.\ C {\bf 31}, 1 (2003)
  [arXiv:hep-ex/0210014];
  J.~Abdallah {\it et al.}  [DELPHI Collaboration],
  Eur.\ Phys.\ J.\ C {\bf 36}, 1 (2004)
  [Eur.\ Phys.\ J.\ C {\bf 37}, 129 (2004)]
  [arXiv:hep-ex/0406009].

\bibitem{Janot:2004cy}
  See, for example, P.~Janot,
  Phys.\ Lett.\ B {\bf 594}, 23 (2004)
  [arXiv:hep-ph/0403157].

\bibitem{Carpenter:2006hs}
  L.~M.~Carpenter, D.~E.~Kaplan and E.~J.~Rhee,
  arXiv:hep-ph/0607204.

\bibitem{AandC}
  See their websites:  cern.ch/atlas
  and cms.cern.ch/

\bibitem{Baer:1996wa}
  H.~Baer, C.~h.~Chen and X.~Tata,
  Phys.\ Rev.\  D {\bf 55}, 1466 (1997)
  [arXiv:hep-ph/9608221].
\bibitem{Prescale}
   See websites\\ twiki.cern.ch/twiki/bin/view/CMS/OnSel\_01\_V\_06\\ and \\
   lxmon02.cern.ch/twiki/bin/view/Atlas/JetTiggerTable .

\bibitem{ReOpt}
  LHCb Collaboration,
  ``Status of LHCb Detector Reoptimization'',
  CERN/LHCC 2003-003 (2003)

\bibitem{Hinchliffe:1992ad}
  I.~Hinchliffe and T.~Kaeding,
  Phys.\ Rev.\  D {\bf 47}, 279 (1993).
%

\bibitem{schneider}
  O. Schneider \textit{Private Communication}.

\bibitem{Muheim:Beauty06}
 F.~Muheim, LHCb Collaboration,
 hep-ex/0703006v1 (2007)


\bibitem{kirill} C. Anastasiou $\&$ K. Melnikov hep-ph/0207004

\bibitem{Rparity} Linda M. Carpenter, David E. Kaplan, Eun-Jung Rhee hep-ph/0607204

\bibitem{thesis} C. Currat CERN-THESIS-2001-024 (2001)


\bibitem{Alpgen} M. L. Mangano,M. Moretti, F. Piccinini, R. Pittau and A. D. Polosa, JHEP 0307, 001 (2003), hep-ph/0206293.

\bibitem{nakada} T. Nakada, Private correspondence (2007)

\bibitem{petar} P. Maksimovic, Private correspondence (2007)


\end{thebibliography}
\end{document}